# Glass stability (GS) of chemically complex (natural) sub-alkaline glasses

G. Iezzi[1,2], A.L. Elbrecht[3], M. Davis[3], F. Vetere[4],
V. Misiti[2], S. Mollo[5,2], A. Cavallo[6,2]

[1]Dep. "Ingegneria & Geologia (InGeo), University G. d'Annunzio of Chieti-Pescara, Italy;
[2]SCHOTT North America, New York, USA;
[3]Department of Physics and Geology University of Perugia Italy;
[4]INGV Rome Italy;
[5]Dep. "Scienze della Terra", University La Sapienza University of Roma, Italy.
[6]CETREMA Grosseto, Italy;

Corresponding author: Gianluca Iezzi (g.iezzi@unich.it)

Keywords: DSC, SEM, EPMA, GS (glass stability), GFA (glass-forming ability).

**Abstract**

Glass stability (GS) indicates the glass reluctance or ability to crystallise upon heating; it can be characterised by several methods and parameters and is frequently used to retrieve glass-forming ability (GFA) of corresponding liquids as the case with which such liquids can be made crystal free via melt-quenching. Here, GS has been determined for the first time on six sub-alkaline glasses having complex (natural) compositions, the most widespread and abundant on Earth. The two end-members are a basalt and a rhyolite, $B_{100}$ and $R_{100}$, plus intermediate compounds $B_{80}R_{20}$, $B_{60}R_{40}$, $B_{40}R_{60}$, $B_{20}R_{80}$. Each glass was heated in a differential scanning calorimetry (DSC) at a rate of 10 °C/min (600 °C/h) to measure *in-situ* $T$g (glass transition), $T$x (onset of crystallization) and $T$m (melting) temperatures, from ambient to their *liquidus* temperatures. The *ex-situ* run-products quenched at $T$m have been characterised by SEM and EPMA to quantify textures and compositions of phases, respectively.

$R_{100}$ and $B_{20}R_{80}$ do not shown any DSC peak, whereas $B_{40}R_{60}$, $B_{60}R_{40}$, $B_{80}R_{20}$ and $B_{100}$ thermograms display progressively more resolvable peaks. As $SiO_2$ (wt.%) in the system increases from $B_{100}$ to $B_{40}R_{60}$, $T$x linearly increases, $T$m first decreases and then levels off, whereas $T$g poorly changes. In agreement, $R_{100}$ and $B_{20}R_{80}$ run-products are completely glassy, while from $B_{100}$ to $B_{40}R_{60}$ the amount of glass (gl) increases from 48.5 to 97 area%, counterbalanced by a decreasing of clinopyroxene (cpx) from 47.7 to 16 area%, whereas spinel (sp) accounts of only 0.9 to 3.8 area%. Plagioclase (plg) crystallises heterogeneously on the $Al_2O_3$ holders only in $B_{100}$ and $B_{80}R_{20}$ and for distance < 100 μm from it. $R_{100}$, $B_{20}R_{80}$, $B_{40}R_{60}$ and $B_{60}R_{40}$ *ex-situ* glasses have chemistries very close to their starting compositions, according to the absence or scarcity of crystals formed during heating. Instead, $B_{100}$ and $B_{80}R_{20}$ glasses are enriched in Si, Al and Na and depleted in Fe, Mg and Ca due to internal crystallization of sp and mostly cpx. Cpx in $B_{100}$ is rich in $^{M2}Ca$, $^{M1}Mg$, $^{M2,M1}Fe$ and remarkably of $^{M1,T}Al$.

$K_T$, $K_H$, $K_W$, $K_{LL}$ and $w_2$ GS parameters increase linearly and monotonically as a function of $SiO_2$, with very high correlations. Moreover, $T$x values and GS parameters highly correlate with GFA *via* $R$c (critical cooling rate), previously determined with *ex-situ* cooling-induced experiments. Therefore, GS scales with GFA for natural silicate compositions. In addition, the *in-situ* $R$c value of $B_{100}$ measured with DSC results > 45 °C/min (> 2700 °C/h), broadly corroborating the $R$c of about 150 °C/min (9000 °C/h) determined *ex-situ*. In turn, relevant solidification parameters on heating or cooling can be obtained by DSC investigations also for chemically complex (natural) systems, similar to simple silicate systems. These outcomes are relevant for lavas or magmas that re-heat glass-bearing volcanic rocks, as well as for fabricate glass-ceramic materials with desirable texture and composition of phases starting from abundant and very cheap raw volcanic rocks.

## Introduction

In Earth Sciences, dynamic solidification processes have been studied mainly by *ex-situ* cooling-induced ([1-11] and reference therein) and to a lesser extent by decompression-induced degassing experiments [4, 12-15], starting from a dry or volatile-bearing silicate liquid ± crystals. The major part of these studies focus on $SiO_2$-poor systems, relevant for basaltic lavas and magmas, whereas more $SiO_2$-rich systems are by far lesser investigated [4, 6, 10-11]. Even fewer investigations are available for solidification behaviours induced by heating or re-heating on glasses, except for a few studies on basaltic to basaltic-andesite systems [16-22]. In parallel, differential scanning calorimetry (DSC) and/or differential thermal analysis (DTA) techniques are rarely applied on crystallization of liquids and glasses relevant to geological systems, although these *in-situ* methods are useful and rapid to quantify kinetic parameters [23-24].

This contrasts with common using of such *in-situ* approaches in Glass Sciences, especially for characterizing solidification of non-crystalline solids upon heating ([25-37] and references therein). DSC is commonly used to retrieve the melting ($T$m) and glass transitions ($T$g) temperatures, plus key $T$x (or $T$c) values, i.e. the temperature of onset of crystallization. To the best of our knowledge, only [24, 38] provided crucial $T$x determinations for basaltic glasses or, in general, for non-crystalline solids with chemically complex compositions as the natural ones. Most systems studied in glass sciences focus on simple compounds with only few (mainly two or three, rarely four) major components; typically, these systems solidify phases on heating or cooling, with compositions identical to those of parent glasses and liquids, respectively [23]. Instead, in natural silicate systems the number of major components are normally seven to eight and solidified phases which have stoichiometry very different from the parent liquid on cooling and/or glass during heating.

From the above literature studies is clear that there is a need to further study nucleation and growth kinetic in order to shed new light on one of the most important phenomenon linked not only to magmatic/volcanic events but also to industry. Therefore, we present new DSC investigations upon heating on six chemical complex systems. Experiments were performed by using a ramp rate of 10 °C/min from room to melting conditions. These six sub-alkaline natural compounds have chemistries ranging from $SiO_2$-poor basalt ($B_{100}$) to $SiO_2$-rich rhyolite ($R_{100}$), for which GFA is already known and determined by cooling-induced experiments via the so called *critical cooling rate* ($R$c) method (for details refer to [10-11]). Experiments on re-heated and then quenched (from $T$m) charges have been also structurally and chemically characterized by SEM and EPMA, to corroborate DSC data and to quantify textures and compositions of all produced phases. The attained outcomes can be useful in scenarios where glass-bearing rocks are re-heated, i.e. flowage of lavas on volcanic rocks [16-19].

In parallel, the huge abundance of these compounds, their very cheap costs and the less effort for preparation could represent valuable alternatives to design new glass-ceramics with desired macroscopic properties for encapsulation of toxic substances.

### Starting glasses, experimental and analytical methods

*Starting glasses*. The six different silicate starting glass compositions are the same used in [10-11]. Briefly, they were prepared by using two natural rocks: a basalt from Iceland ($B_{100}$) and rhyolite from the Lipari Island in the Aeolian arc ($R_{100}$). About 100 g of these two rocks were powdered and melted in Pt crucibles at temperature of 1600 °C for 4 h in air. After quenching on metal plates, they were crushed and re-melted two times at the same conditions in order to improve homogeneity. Then, four intermediate compositions were obtained by mechanically mixing $B_{100}$ and $R_{100}$ powders in the following proportions (by wt.%): $B_{80}R_{20}$, $B_{60}R_{40}$, $B_{40}R_{60}$, $B_{20}R_{80}$. These glassy powders were, afterwards, heat treated at identical conditions of the two end-members. The final six starting glassy compositions with systematic chemical variations, measured by EPMA (see below), are reported in Tab. 1, where are also reported the amounts of dissolved $H_2O$ (measured by FTIR technique), densities, $Fe^{2+}/Fe_{tot}$ ratios and corresponding nomenclatures in the TAS classification diagram [10-11].

*DSC (Differential Scanning Calorimetry)*. Single piece each ca. 125 mg of $B_{100}$, $B_{80}R_{20}$, $B_{60}R_{40}$, $B_{40}R_{60}$, $B_{20}R_{80}$ and $R_{100}$ starting glass were loaded in $Al_2O_3$ crucibles. The sample charges were then lodged in the DSC (Netzsch STA 449 C) and run with a heating rate of 10 °C/min, from room to melting conditions (1280 °C for of $B_{100}$ to 1130 °C for $R_{100}$), previously estimated by thermodynamic models [10]. When the liquidus region for each system was achieved, the sample-charges were rapidly quenched to ambient conditions. A duplicate DSC experiment was run for $B_{100}$ using the same thermal program, but instead a Pt-crucible was used in order to evaluate the possible effect that sample holder could have on phase nucleation and growth. The same apparatus was also used to bracket *in-situ* the $R$c (GFA) of $B_{100}$ [10, 11] following the method of [38].

*SEM and EPMA analyses*. The quenched run products were mounted in epoxy and polished for textural and chemical analysis. Images of phases and their textures were first identified and collected by back-scattered electrons using a field-emission JEOL 6500 F scanning electron microscopy, equipped with an energy-dispersive spectrometer (EDS). Their chemical attributes were then accurately determined with a JEOL JXA 8200 electron probe micro-analyzer equipped with five wavelength-dispersive spectrometers (WDS). Both these facilities are installed at the HP-

HT laboratory of Volcanology and Geophysics of Istituto Nazionale di Geofisica e Vulcanologia (Rome, Italy). The operative conditions and data reduction obtained by FE-SEM and EPMA are extensively reported in [39-41] and [10-11].

In particular, the chemical characterization of phases with a diameter > 1 and < 3 μm was obtained by the EDS of SEM, whereas those ≥ 3 μm by EPMA. The differences of average chemical compositions measured either by EPMA and FE-SEM were negligible, as indicated by analyses on the same glass matrix portions as reported in [11]. The textures of run-products, i.e. type, size and distribution of each phase, have been first evaluated on back-scattered SEM microphotographs collected at 150, 400, 800 and 1500 X magnifications. As a function of these first qualitative observations, image analysis has been performed following classical procedures reported in details by [3, 39-43]. Such analyses have been achieved on BSE images that better represent the whole texture in all the run-products. Quantitative textures of crystal-bearing $B_{100}$, $B_{80}R_{20}$ and $B_{60}R_{40}$ run-products were considered in two ways, i.e. in contact and beyond a distance of 100 μm from the $Al_2O_3$ sample holder wall in order to evaluate possible effects of $Al_2O_3$ on nucleation [10, 44].

## Results

*DSC*. The six DSC spectra for $B_{100}$, $B_{80}R_{20}$, $B_{60}R_{40}$, $B_{40}R_{60}$, $B_{20}R_{80}$ and $R_{100}$ collected from experiments performed in $Al_2O_3$ crucibles and those recorded in Pt for $B_{100}$, are stacked in Fig. 1. $R_{100}$ and $B_{20}R_{80}$ DSC spectra do not show any evident peaks. $T$g and $T$m peaks for $B_{100}$, $B_{80}R_{20}$, $B_{60}R_{40}$ and $B_{40}R_{60}$ have respectively very low to high endothermic intensities, whereas $T$x is marked by high exotermic peaks. $T$g, $T$x and $T$m are progressive less noticeable when moving from $SiO_2$-poor ($B_{100}$) to $SiO_2$-rich ($B_{40}R_{60}$) systems (Fig. 1). The DSC thermograms of $B_{100}$ run with $Al_2O_3$ and Pt sample holders show identical $T$g, $T$x and $T$m peak positions, but the ratio between signal and background is higher for spectrum collected with the Pt charge than the $Al_2O_3$; furthermore, only the DSC spectra of $B_{100}$ show a low intense exothermic peak between $T$x and $T$m, indicative of a secondary crystallization event (Fig. 1).

The values of $T$g, $T$x and $T$m are reported in Tab. 2. In Tab. 1 are also reported the modelled values of *$T$m and *$T$g, showing differences with $T$m and $T$g measured by DSC respectively of 29 and 41 °C for $B_{100}$, 9 and 29 °C for $B_{80}R_{20}$, 20 and 27 °C for $B_{60}R_{40}$ and of 9 and 18 °C for $B_{40}R_{60}$. Hence, the DSC kinetic determinations of $T$g and $T$m agree with those calculated at equilibrium conditions with available thermodynamic and rheological models, being differences lesser than 30 and 42 °C, respectively. The trend of $T$g, $T$x and $T$m as a function of $SiO_2$ of the bulk system are diagrammed in Fig. 2. $T$g changes of only 10 °C (Tab. 2), $T$m first decreases and then levels off and $T$x show evident higher linear increasing moving from $B_{100}$ to $B_{40}R_{60}$ (Fig. 2).

The GFA character of $B_{100}$ has been measured *in-situ* here, following the approach described in [24, 38] of heating and cooling cycles. The related thermograms and derived peaks areas, as a function of cooling rate, are displayed in Fig. 3. The area of peaks moderately increases for rates between 5 and 10 and then rapidly augments from 10 to 45 K/min; however, the plateau value indicative of the $R$c has been un-attained. Hence, DSC data indicate that the $R$c value for $B_{100}$ is higher than 45 °C/min or 4050 °C/h in agreement with the $R$c of about 9000 °C/h determined by serial cooling rate *ex-situ* experiments by [10-11].

*Textures of phases*. The general and salient textural features of the $B_{100}$, $B_{80}R_{20}$, $B_{60}R_{40}$ and $B_{40}R_{60}$ crystal-bearing run-products are displayed at variable magnifications in Fig. 4. Qualitatively, the increasing of $SiO_2$ in the bulk system ($B_{100}$ to $B_{40}R_{60}$) induces a reduction of the amount and size of crystalline phases. Sp is ubiquitous in any of the run-products with size of several μm in $B_{100}$, around 1 μm in $B_{80}R_{20}$ and sub-micrometric for both $B_{60}R_{40}$ and $B_{40}R_{60}$. Cpx occurs in $B_{100}$, $B_{80}R_{20}$ and $B_{60}R_{40}$ with size of few tens of μm in $B_{100}$ with irregular to dendritic shapes, lengths lower than 10 μm in $B_{80}R_{20}$ mainly with dendritic aspect and like tiny elongated dendrites of only few μm in $B_{60}R_{40}$ system (Fig. 4). Cpx crystals branches out preferentially from single sp crystals; textures of both sp and cpx appears distributing along strips in $B_{100}$, more randomly and homogeneously spread in $B_{80}R_{20}$, whereas in $B_{60}R_{40}$ cpx and sp are agglomerated in relative crystal-rich patches in an intra-crystalline dark glass, surrounded by a glass matrix (Fig. 4). On the other hand, plg is present only on the rim of $Al_2O_3$ sample holder, showing prismatic to acicular crystal shape with the longest size nearly perpendicular to capsule walls (Fig. 4). Plg is relatively abundant in $B_{100}$, moderate for $B_{80}R_{20}$, very low for $B_{60}R_{40}$ and undetectable for the $B_{40}R_{60}$ system; plg disappears at a linear distance from the $Al_2O_3$ wall of some tens of μm in $B_{100}$ and even at shorter distance (closer to the sample holder) for the other two $B_{80}R_{20}$ and $B_{60}R_{40}$ compositions (Fig. 4).

The area% of plg, cpx, sp and glass phases are reported in Tab. 3, either far or close from alumina surfaces; the former situation is unaffected by $Al_2O_3$ and thus is representing the actual and intrinsic crystallization behaviour of these whole glass systems. The evolution of phases contents as a function of bulk $SiO_2$ is shown in Fig. 5. The amount of sp is < 4 area% and follows a moderate concave upward trend with maxima at $B_{100}$ and $B_{40}R_{60}$ system. Cpx area% decrease from about 50 to 40 between $B_{100}$ and $B_{80}R_{20}$, down to 16 area % in $B_{60}R_{40}$. The glass phases, either matrix or intra-crystalline, rapidly and almost linear increasing from $B_{100}$ to $B_{40}R_{60}$ and it represents the unique phase in $B_{20}R_{80}$ and $R_{100}$ (Fig. 5).

*Compositions of phases*. The majority of crystalline phases have very tiny size dimensions; in turn, only cpx crystals in $B_{100}$ can be quantifiable by EPMA (Tab. 4). The chemical features of the major oxide amounts in the starting glass systems and those measured after quenching are compared in Fig. 6. Oxides in $B_{60}R_{40}$, $B_{40}R_{60}$, $B_{20}R_{80}$ and $R_{100}$ quenched glasses have almost identical compositions of their bulk chemistries, whereas both $B_{100}$ and $B_{80}R_{20}$ are enriched in $SiO_2$, $Al_2O_3$ and $Na_2O$ and poorer in $TiO_2$ (only for $B_{100}$), $Fe_2O_3$, MgO and CaO, respectively (Fig. 6). The correspondence of all chemical oxides of $B_{60}R_{40}$, $B_{40}R_{60}$, $B_{20}R_{80}$ and $R_{100}$ on 1:1 line corroborate the very low amount and absence of crystalline phases measured by textures (Tab. 3 and Fig. 5), whereas those quantified in $B_{100}$ and $B_{80}R_{20}$ are mainly due to the solidification of cpx (Tab. 4 and Fig. 6). The average crystal-chemical formula of it in $B_{100}$, according to EPMA data (Tab. 4) and crystal-chemical constrains of single-chain clinopyroxenes [45-46] is $^{M2}(Na_{0.03}Ca_{0.68}Fe^{2+}_{0.18}Mg_{0.11})^{M1}(Mg_{0.67}Fe^{3+}_{0.09}Al_{0.21}Ti_{0.03})^{T}(Al_{0.32}Si_{1.68})O_6$; the marked peculiarity of this cation ordering is the presence of significant amount of Al, either in T- and M1-sites, similarly to cpx grown under kinetics conditions in an alkali basalt by [47].

**Discussion**

The GS attributes of all the six bulk chemical systems investigated by *in-situ* DSC (Tabs. 1 and 2, Figs. 1 and 2) corroborate well by *ex-situ* textural and micro-chemical outcomes (Tabs. 3 and 4, Figs. 3, 4 and 5). The heating rate of 10 °C/min or 600 °C/h is too high to permit crystal nucleation in $R_{100}$ and $B_{20}R_{80}$. It is close to a critical heating rate for $B_{40}R_{60}$ and becomes progressively more amenable to crystal nucleation for $B_{60}R_{40}$, $B_{80}R_{20}$ and $B_{100}$ systems, according to their amounts of crystals (Tabs. 2 and 3, Figs. 1, 2, 3 and 4). Thus, $B_{60}R_{40}$, $B_{80}R_{20}$ and $B_{100}$ chemical systems require progressive higher rates to avoid nucleation upon heating, whereas $R_{100}$ and $B_{20}R_{80}$ necessitate lower heating rates to initiate nucleation. Therefore, the amount of $SiO_2$ or the NBO/T parameter scale with a critical heating rate and GS, similarly to those demonstrated experimentally with GFA *via* $R$c [11].

The thermal range between $T$m and $T$g for $B_{100}$, $B_{80}R_{20}$, $B_{60}R_{40}$ and $B_{40}R_{60}$ is 512, 501, 462 and 460 °C, respectively (Tab. 2). Since the heating rate is 10 °C/min, these glasses were re-heated dynamically for 50/45 minutes between $T$g and $T$m, allowing crystallization of 51.5, 41.5, 17.4 and only 3 area% (Tab. 2) of crystals for $B_{100}$, $B_{80}R_{20}$, $B_{60}R_{40}$ and $B_{40}R_{60}$, respectively. These results globally corroborate with those obtained by isothermal (fixed *T*, variable *t*) re-heating experiments and related TTT diagrams performed on basaltic to basaltic-andesites glasses by [20-22]. These previous studies highlight that re-crystallization is extensive and requires even few minutes at intermediate *T*, whereas is moderate to weak and necessitates tens of minutes when isothermal treatment is close to $T_m$ or $T_g$.

These GS behaviours of chemically-complex sub-alkaline systems can be validated here quantitatively and can be compared and relates with GFA. GS has been computed in different ways, using the most known and used parameters:

$K_T = T_g / T_m$ [48]

$K_H = (T_x - T_g) / (T_m - T_x)$ [22]

$K_W = (T_x - T_g) / T_m$ [23]

$K_{LL} = T_x / (T_g + T_m)$ [26]

$w_2 = (T_g / T_m) - (T_g / (2T_x - T_g))$ [49].

which combine $T_m$, $T_g$ and $T_x$, (Tab. 2). According to DSC data and critical heating rate (see above), in Fig. 7 are displayed all the GS parameters as a function of $SiO_2$, i.e. those of $B_{40}R_{60}$, $B_{60}R_{40}$, $B_{80}R_{20}$ and $B_{100}$ systems. Each GS parameter increases when $SiO_2$ augments, along linear trends. The linear regressions show relatively low $R^2$ for $K_T$, the so-called reduced glass transition, which does not include $T_x$, whereas all the others including $T_x$ have very high $R^2$ values (Tab. 2 and Fig. 7). The same relationships and regressions are provided when GS parameters are plotted *versus* NBO/T (Tab. 1). Therefore, $T_x$ and GS are both linearly proportional to sub-alkaline chemical composition, either $SiO_2$ or NBO/T (Figs. 2 and 7); in other words, enrichment of $SiO_2$ or decreasing of NBO/T in sub-alkaline glasses enhance their stability upon heating or, alternatively, progressively impede their nucleation.

The GS attributes are frequently used to retrieve the GFA [26-37, 48-49]. Therefore, it is possible to relate GS parameters (by considering those glass as the most widespread compositions on Earth), by comparing Rc with GS parameters as well as with $T_x$. These comparisons are displayed in Fig. 8, showing that the increasing of *R*c as a function of bulk chemical systems for $B_{100}$, $B_{80}R_{20}$, $B_{60}R_{40}$, and $B_{40}R_{60}$ (determined in [11]), is inversely and linearly correlated with *Tx* and $K_W$, $K_{LL}$ and $w_2$, with high correlations degree (Fig. 8).

These relationships prove that natural sub-alkaline glasses, characterised by a large amount of chemical elements (O, Si, Ti, Al, Fe, Mg, Ca, Na and K), remain more easily in glassy state when heated from ambient to *liquidus* conditions and more easily will form a glass when cooled from *liquidus* to room conditions as a function of increasing $SiO_2$ or decreasing of NBO/T. The straightforward corollary is that the decreasing of $SiO_2$, or increasing of NBO/T, progressively facilitates nucleation. Under natural conditions, $SiO_2$-rich glasses hosted in volcanic rocks have high chance to persist to thermal perturbations, whereas $SiO_2$-poor ones can be transformed relatively fast and easily, being prone to nucleate. The same general behaviour holds under cooling from completely molten bulk silicate systems. The GS and GFA characteristics highlighted here, as a function of compositions and thermal treatments, can be used to design glasses and glass-ceramics with the lowest costs of raw-materials and by far the most abundant on Earth.

## Acknowledgments

This work was funded by the "fondi ateneo" of the University G. d'Annunzio of Chieti-Pescara assigned to Iezzi G.. Vetere F. acknowledges the European Research Council for the Consolidator Grant ERC-2013-CoG Proposal No. 612776 — CHRONOS to Perugini D. and support from the TESLA FRB-base project from the Department of Physics and Geology, University of Perugia.

# References

[1] K.V. Cashman, Relationship between plagioclase crystallization and cooling rate in basaltic melts Contributions to Mineralogy and Petrology 113 (1989) 126-142.

[2] A.C. Lasaga, Kinetic Theory in the Earth Sciences. Princeton University Press, Princeton, New York (1997).

[3] M.D. Higgins, Quantitative Textural Measurements in Igneous and Metamorphic Petrology. Cambridge University Press, Cambridge (2006).

[4] J.E. Hammer, Experimental studies of the kinetics and energetics of magma crystallization, in: Putirka, K.D., Tepley, F.J. (Eds.). Reviews in Mineralogy 69 (2008) 9-59.

[5] G. Iezzi, S. Mollo, G. Ventura, A. Cavallo, C. Romano, Experimental solidification of anhydrous latitic and trachytic melts at different cooling rates: the role of nucleation kinetics. Chemical Geology 253 (2008) 91-101.

[6] G. Iezzi, S. Mollo, G. Ventura, Solidification behaviour of natural silicate melts and volcanological implications, in: Lewis, N., Moretti, A. (Eds.), Volcanoes: Formation, Eruptions and Modelling. Nova publishers, New York (2009) 127-151.

[7] G. Iezzi, S. Mollo, Torresi, G. Ventura, A. Cavallo, P. Scarlato, Experimental solidification of an andesitic melt by cooling. Chemical Geology 283 (2011) 261-273.

[8] P. Del Gaudio, S. Mollo, G. Ventura, G. Iezzi, J. Taddeucci, A. Cavallo, Cooling rate-induced differentiation in anhydrous and hydrous basalts at 500 MPa: implications for the storage and transport of magmas in dikes. Chemical Geology 270 (2010) 164-178.

[9] S. Mollo, G. Iezzi, G. Ventura, A. Cavallo, P. Scarlato, Heterogeneous nucleation mechanisms and formation of metastable phase assemblages induced by different crystalline seeds Journal of Non-Crystalline Solids 358 (2012) 1624-1628.

[10] F. Vetere, G. Iezzi, H. Behrens, A. Cavallo, V. Misiti, M. Dietrich, J. Knipping, G. Ventura, S. Mollo, Intrinsic solidification behaviour of basaltic to rhyolitic melts: a cooling rate experimental study. Chemical Geology 354 (2013) 233-242.

[11] F. Vetere, G. Iezzi, H. Behrens, F. Holtz, G. Ventura, V. Misiti, A. Cavallo, S. Mollo, M. Dietrich, Glass forming ability and crystallization behaviour of sub-alkaline silicate melts. Earth-Science Reviews 150 (2015) 25-44.

[12] P. Armienti, Decryption of igneous rock textures: crystal size distribution tools, in: Putirka, K.D., Tepley, F.J. (Eds.). Reviews in Mineralogy 69 (2008) 623-649.

[13] L.J. Applergarth, H., Tuffen M.R. James, H. Pinkerton, Earth-Science Reviews 116 (2013) 1-16.

[14] Arzilli, F., Carroll, M.R. (2013). Crystallization kinetics of alkali feldspars in cooling and decompression-induced crystallization experiments in trachytic melt. Contributions to Mineralogy and Petrology 166 1011-1027.

[15] A. Fiege, F. Vetere, G. Iezzi, A. Simon, F. Holtz, The roles of decompression rate and volatiles ($H_2O+Cl\pm CO_2\pm S$) on crystallization in (trachy-) basaltic magma. Chemical Geology 411 (2015) 310-322.

[16] D.J.M. Burkhard, Crystallization and oxidation during emplacement of lava lobes. Special Paper of the Geological Society of America 396 (2005a) 67-80.

[17] D.J.M. Burkhard,. Nucleation and growth rates of pyroxene, plagioclase, and Fe-Ti oxides in basalt under atmospheric conditions. European Journal of Mineralogy 17 (2005b) 675-685.

[18] D.J.M. Burkhard, Thermal interaction between lava lobes. Bulletin of Volcanology 65 (2-3) (2003) 136-143.

[19] D.J.M. Burkhard, Crystallization and oxidation of Kilauea basalt glass: Processes during reheating experiments. Journal of Petrology 42 (3) (2001) 507-527.

[20] N. Deardor, K. Cashman, Rapid crystallization during recycling of basaltic andesite tephra: timescales determined by reheating experiments. Scientific Reports 7 (2017) 46364

[21]. D'Oriano, C., M. Pompilio, A. Bertagnini, R. Cioni, M. Pichavant, Effects of experimental reheating of natural basaltic ash at different temperatures and redox conditions. Contrib Mineral Petr 165 (2013) 863–883 doi: 10.1007/s00410-012-0839-0.

[22]. C. D'Oriano, A. Bertagnini, R. Cioni, M. Pompilio, Identifying recycled ash in basaltic eruptions. Sci Rep-Uk 4, (2014) doi: ARTN 585110.1038/srep05851.

[23] J.E. Shelby, Introduction to Glass Science and Technology, 2nd edition. Padstow, Conwall, UK (2005).

[24] C.S. Ray, S.T. Reis, S. Sen, J.S. O'Dell, JSC-1A lunar soil simulant: Characterization, glass formation, and selected glass properties. Journal of Non-Crystalline Solids 356 (2010) 2369-2374.

[25] A. Hrubÿ, Evaluation of glass-forming tendency by means of DTA. Czechoslovak Journal of Physics B 22 (1972) 1187–1193.

[26] M.C. Weinberg, Glass-forming ability and glass stability in simple systems. Journal of Non-Crystalline Solids 167 (1994) 81-88.

[27] A.A. Cabral, C. Fredericci, E.D. Zanotto, A test of the Hrubÿ parameter to estimate glass-forming ability. Journal of Non-Crystalline Solids 219 (1997) 182-186.

[28] A.A. Cabral, A.A.D. Cardoso, E.D. Zanotto, Glass-forming ability versus stability of silicate glasses. I. Experimental test. Journal of Non-Crystalline Solids 320 (2003) 1-8

[29] Z.P. Lu, C.T. Liu, A new glass-forming ability criterion for bulk metallic glasses. Acta Mater. 50 (2002) 3501-3512.

[30] Long, Z., Wei, H., Ding, Y., (...),Xie, G., Inoue, A. (2009). A new criterion for predicting the glass-forming ability of bulk metallic glasses. Journal of Alloys and Compounds, 475(1-2), pp. 207-219.

[31] Zhang, P., Wei, H., Wei, X., Long, Z., Su, X. (2009). Evaluation of glass-forming ability for bulk metallic glasses based on characteristic temperatures. Journal of Non-Crystalline Solids, 355, 2183-2189.

[32] Nascimento, M.L.F., Souza, L.A.,Ferreira, E.B., Zanotto, E.D. (2005). Can glass stability parameters infer glass forming ability?. Journal of Non-Crystalline Solids, 351(40-42), pp. 3296-3308.

[33] A.F. Kozmidis-Petrović, Sensitivity of the Hruby, Lu-Liu, Fan, Yuan, and Long glass stability parameters to the change of the ratios of characteristic temperatures $T_x/T_g$ and $T_m/T_g$. Thermochimica Acta 510 (2010a) 137-143.

[34] A.F. Kozmidis-Petrović, Theoretical analysis of relative changes of the Hruby, Weinberg, and Lu-Liu glass stability parameters with application on some oxide and chalcogenide glasses. Thermochimica Acta 510 (2010b) 54-60

[35] A.F. Kozmidis-Petrović, Which glass stability criterion is the best? Thermochimica Acta 523 (2011) 116-123

[36] E. B. Ferreira, E. D. Zanotto, S. Feller, G. Lodden, J. Banerjee, T. Edwards,M. Affatigato, Critical Analysis of Glass Stability Parameters and Application to Lithium Borate Glasses 94 (2011) 3833-3841

[37] T.V.R. Marques, A.A. Cabral, Influence of the heating rates on the correlation between glass-forming ability (GFA) and glass stability (GS) parameters. Journal of Non-Crystalline Solids 390 (2014) 70-76.

[38] C.S. Ray, S.T. Reis, R.K. Brow, W. Höland, V. Rheinberger, A new DTA method for measuring critical cooling rate for glass formation. Journal of Non-Crystalline Solids 351 (2005) 1350-1358.

[39] G. Iezzi, S. Mollo, G. Ventura, A. Cavallo, C. Romano, Experimental solidification of anhydrous latitic and trachytic melts at different cooling rates: the role of nucleation kinetics. Chem. Geol. 253 (2008) 91–101.

[40] G. Iezzi, S. Mollo, G. Torresi, G. Ventura, A. Cavallo, P. Scarlato, Experimental solidification of an andesitic melt by cooling. Chem. Geol. 283 (2011) 261–273.

[41] G. Iezzi, S. Mollo, E. Shahini, A. Cavallo, P. Scarlato, The cooling kinetics of plagioclase feldspars as revealed by electron-microprobe mapping. American Mineralogist 99 (2014a) 898–907.

[42] G. Iezzi, C. Caso, G. Ventura, M. Vallefuoco, A. Cavallo, H. Behrens, S. Mollo, D. Paltrinieri, P. Signanini, F. Vetere, Historical volcanism at the Marsili Seamount (Tyrrhenian Sea, Italy): first documented deep submarine explosive eruptions in the Mediterranean Sea. Gondwana research 25 (2014b) 764-774.

[43] G. Lanzafame, S. Mollo, G. Iezzi, C. Ferlito, G. Ventura, Unraveling the solidification path of a pahoehoe "cicirara" lava from Mount Etna volcano. Bulletin of Volcanology 75 (2013) 1-16.

[44] S. Mollo, G. Iezzi, G. Ventura, A. Cavallo, P. Scarlato Heterogeneous nucleation mechanisms and formation of metastable phase assemblages induced by different crystalline seeds in a rapidly cooled andesitic melt. Journal of Non-Crystalline Solids 358 (2012)1624-1628.

[45] W.A Deer, R.A. Howie, and J. Zussman, Rock forming minerals, Doublechain Silicates. Longman Scientific and Technical (1997).

[46] G. Iezzi, G. D. Bromiley, A. Cavallo, P. P. Das, F. Karavassili, I. Margiolaki, A. A. Stewart, M. Tribaudino, J. P. Wright, Solid solution along the synthetic $LiAlSi_2O_6$-$LiFeSi_2O_6$ (spodumene-ferri-spodumene) join: A general picture of solid solutions, bond lengths, lattice strains, steric effects, symmetries, and chemical compositions of Li clinopyroxenes, American Mineralogist 101 (2016) 2498-2513.

[47] S. Mollo, P. Del Gaudio, G. Ventura, G. Iezzi, P. Scarlato, Dependence of clinopyroxene composition on cooling rate in basaltic magmas: implications for thermobarometry. Lithos, 118 (2010) 302-312.

[48] D. Turnbull, Under what conditions can a glass be formed? Contemporary Physics 10 (1969) 473-488.

[49] B. Gu, F. Liu, Y. Jiang, K. Zhang, Evaluation of glass-forming ability criterion from phase-transformation kinetics. Journal of Non-Crystalline Solids 358 (2012) 1764-1771.

[50] G.H. Klöß, Dichtefluktuationen natürlicher Gläser (Dissertation) University of Jena (2000).

[51] H. Hui, Y. Zhang, Toward a general viscosity equation for natural anhydrous and hydrous silicate melts. Geochim. Cosmochim. Acta 71 (2007) 403–416.

**Table captions**

**Table 1**. Chemical compositions of the six glass starting materials.

Footnotes: averages and standard deviations of major components are in wt. %, obtained on at least 15 EPMA point analyses; $Fe^{2+}/Fe_{tot}$ was determined for each sample using a modified Wilson method as described in [10]. $H_2O$ content was measured by IR spectroscopy using the peak height of the absorption band at 3550 cm$^{-1}$ [10]. Glass density was estimated using the model of [50] already reported in [10]. $T$m and $T$g (°C at 10$^{12}$ Pa s) are calculated using MELTS (Ghiorso and Sacks, 1994) and the general viscosity model from [51], respectively. $T$rg is $T$g/$T$m.

**Table 2.** $T_g$, $T_x$ and $T_m$ determined by DSC at a heating rate of 10 °C/min up to 1300 °C from DSC spectra of Fig. 1. The calculated glass stability parameters are also reported.

Footnotes: $K_T = T_g / T_m$ [48]; $K_H = (T_x - T_g) / (T_m - T_x)$ [22]; $K_W = (T_x - T_g) / T_m$ [23]; $K_{LL} = T_x / (T_g + T_m)$ [26]; $w_2 = (T_g / T_m) - (T_g / (2T_x - T_g))$ [49].

**Table 3.** Textural features of run-products quenched after the heating rate of 10 °C/min up to 1300 °C.

Footnotes: *the crystal texture distributions are invariably homogeneous following the criteria proposed by Vetere et al . (2013 and 2014). $B_{100}$ magnifications used 150, 400, 800 and 1500 X; $B_{80}R_{20}$ magnifications used 400, 800 and 1500 X; $B_{60}R_{40}$ magnifications used 800 and 1500 X; $B_{40}R_{60}$ magnifications used 800, 1500 and 4000 X.

**Table 4.** Average and standard deviation chemical compositions of phases in run-products quenched after the heating rate of 10 °C/min up to 1300 °C.

**Figure Captions**

**Figure 1.** DSC spectra of the six glasses heated at 10 °C/min. All glasses were measured using alumina holders, except the $B_{100}$ measured either in Pt (dashed red line) and alumina sample holders. The glass transition, onset of crystallization and melting temperatures ($T_g$, $T_x$ and $T_m$, respectively) are indicated by arrows. The $B_{20}$-$R_{80}$ and $R_{100}$ compositions do not shown any peak. From $B_{100}$ down to $B_{40}R_{60}$ the peaks are progressively less prominent.

**Figure 2.** Evolution of $T_g$, $T_x$ and $T_m$ as a function of $SiO_2$ (wt.%). $T_g$, is nearly constant, whereas $T_x$ and $T_m$ increases and decreases, respectively from $B_{100}$ to $B_{40}R_{60}$.

**Figure 3.** (left) Glass forming ability of $B_{100}$ determined by DSC spectra at various heating and cooling rates following the method of Ray et al. (2005) and (right) related DSC peak area *vs* cooling rate, showing the absence of a plateau indicative that the critical cooling rate ($Rc$) of $B_{100}$ is > 45 °C/min.

**Figure 4.** Phase assemblages by FE-SEM images of run-products heated at 10 °C/min in alumina sample-holders. From $B_{100}$ (top row) to $B_{40}R_{60}$ (last row) the amount of glass (gl) increases, whereas that of pyroxene (px) decreases; the amount of spinel (sp) is invariably few area%. Plagioclase crystallizes heterogeneously only on the alumina sample holders in $B_{100}$ and $B_{80}R_{20}$.

**Figure 5.** Phase amounts as a function of $SiO_2$ (wt.%) of the bulk system. From $B_{100}$ to $B_{40}R_{60}$ the amount of glass (gl) increases, that of pyroxene (px) decreases ($B_{100}$ to $B_{60}R_{40}$) and then disappears ($B_{40}R_{60}$), whereas the amount of spinel (sp) is invariably few area%.

**Figure 6.** Starting chemical compositions *vs* their corresponding glass (residual melt) and cpx (only $B_{100}$ cpx crystals were measurable) compositions. $R_{100}$, $B_{20}R_{80}$, $B_{40}R_{60}$ and $B_{60}R_{40}$ glasses have oxides equal to their starting compositions (except slight differences for $Al_2O_3$ amounts) in agreement with the absence ($R_{100}$, $B_{20}R_{80}$ and $B_{40}R_{60}$) or very slight crystallisation during heating. $B_{100}$ and $B_{80}R_{20}$ have residual melts enriched in $SiO_2$, $Al_2O_3$ and $Na_2O$ and depauperated in $Fe_2O_3$, MgO and CaO compared to their starting oxide corresponding amounts due to a significant crystallisation of sp (spinel) and especially cpx, in agreement with the textural data reported in Table 3 and Figs. 3 and 4. The cpx in $B_{100}$ is relatively rich in CaO and MgO.

**Figure 7**. Principal glass stability parameters (see Table 2 for labels and calculations). All parameters increase linearly and monotonically as a function of $SiO_2$ or NBO/T (see Table 1). The dotted lines are guide for eye, whereas the linear regressions $R^2$ index shown in the legend refer to linear regressions of $SiO_2$ *vs* glass stability parameters.

**Figure 8**. Relations among the critical cooling rates ($R_c$) and their corresponding glass stability parameters (see Table 2 for labels and calculations) for $B_{100}$, $B_{80}R_{20}$, $B_{60}R_{40}$ and $B_{40}R_{60}$ bulk systems, respectively (bottom diagram); relation between the critical cooling rates (Rc) and $T_x$ for $B_{100}$, $B_{80}R_{20}$, $B_{60}R_{40}$ and $B_{40}R_{60}$, respectively (top diagram). Lines are linear regressions with $R^2$ index reported in the legend. The critical cooling rates (Rc) are measured in Vetere et al. (2015). These plots highlights that glass stability and glass forming ability features are strongly related for natural sub-alkaline silicate melts, i.e. a melt and its corresponding glass are both reluctant to nucleate on cooling and heating, respectively.

**Table 1**

| system | SiO$_2$ | TiO$_2$ | Al$_2$O$_3$ | Fe$_2$O$_3$ | MnO | MgO | CaO | Na$_2$O | K$_2$O | P$_2$O$_5$ | total | H$_2$O (ppm) | Fe$^{2+}$/Fe$_{tot}$ | density (g/cm$^3$) | *$T$m (°C) | *$T$g (°C) | $T_{rg}$ | NBO/T | TAS |
|---|---|---|---|---|---|---|---|---|---|---|---|---|---|---|---|---|---|---|---|
| B$_{100}$ | 48.02 (0.40) | 0.98 (0.08) | 15.59 (0.19) | 11.37 (0.25) | 0.18 (0.04) | 9.42 (0.11) | 13.20 (0.14) | 1.79 (0.05) | 0.04 (0.01) | 0.06 (0.02) | 100.65 (0.67) | 53 (13) | 0.386 (0.02) | 2.757 | 1233 | 651 | 0.53 | 0.46 | basalt |
| B$_{80}$R$_{20}$ | 53.01 (0.30) | 0.80 (0.06) | 14.99 (0.12) | 9.49 (0.17) | 0.15 (0.04) | 7.58 (0.10) | 10.79 (0.16) | 2.18 (0.07) | 1.02 (0.03) | 0.02 (0.02) | 100.04 (0.36) | 157 (7) | 0.449 (0.02) | 2.672 | 1193 | 654 | 0.55 | 0.36 | basaltic andesite |
| B$_{60}$R$_{40}$ | 57.97 (0.44) | 0.65 (0.06) | 14.62 (0.27) | 7.73 (0.16) | 0.13 (0.03) | 5.81 (0.08) | 8.46 (0.13) | 2.59 (0.07) | 1.99 (0.05) | 0.04 (0.02) | 99.99 (0.46) | 204 (8) | 0.434 (0.02) | 2.600 | 1170 | 661 | 0.57 | 0.23 | andesite |
| B$_{40}$R$_{60}$ | 62.73 (0.54) | 0.46 (0.05) | 14.05 (0.17) | 6.02 (0.15) | 0.12 (0.04) | 4.01 (0.05) | 6.07 (0.14) | 2.95 (0.07) | 3.02 (0.04) | 0.02 (0.02) | 99.45 (0.73) | 238 (4) | 0.412 (0.02) | 2.525 | 1144 | 675 | 0.60 | 0.17 | andesite |
| B$_{20}$R$_{80}$ | 67.91 (0.38) | 0.29 (0.05) | 13.59 (0.21) | 4.10 (0.20) | 0.11 (0.03) | 2.18 (0.09) | 3.63 (0.13) | 3.29 (0.07) | 3.99 (0.06) | 0.02 (0.02) | 99.10 (0.36) | 384 (20) | 0.424 (0.04) | 2.454 | 1097 | 692 | 0.64 | 0.08 | dacite |
| R$_{100}$ | 73.97 (0.67) | 0.12 (0.06) | 13.48 (0.17) | 2.29 (0.16) | 0.08 (0.05) | 0.44 (0.05) | 1.36 (0.08) | 3.75 (0.17) | 4.89 (0.08) | 0.03 (0.02) | 100.41 (0.84) | 250 (85) | 0.342 (0.02) | 2.368 | 1038 | 732 | 0.69 | 0.01 | rhyolite |

**Table 2**

| system | SiO$_2$ | $T_g$ (°C) | $T_x$ (°C) | $T_m$ (°C) | $K_T$ | $K_H$ | $K_W$ | $K_{LL}$ | $w_2$ |
|---|---|---|---|---|---|---|---|---|---|
| B$_{100}$ | 48.02 | 692 | 905 - 1070 | 1204 | 0.5748 | 0.7124 | 0.1769 | 0.4773 | -0.0442 |
| B$_{80}$R$_{20}$ | 53.01 | 683 | 929 | 1184 | 0.5769 | 0.9647 | 0.2078 | 0.4976 | -0.0044 |
| B$_{60}$R$_{40}$ | 57.97 | 688 | ~ 957 | 1150 | 0.5982 | 1.3938 | 0.2339 | 0.5207 | 0.03709 |
| B$_{40}$R$_{60}$ | 62.73 | 693 | ~ 1000 | 1153 | 0.6010 | 2.0065 | 0.2663 | 0.5417 | 0.07082 |
| B$_{20}$R$_{80}$ | 67.91 | - | - | - | - | - | - | - | - |
| R$_{100}$ | 73.97 | - | - | - | - | - | - | - | - |

**Table 3**

| system | SiO$_2$ | phase contents > 100 μm from Al walls | | | | | phase contents ≤ 100 μm from Al walls | | | | |
|---|---|---|---|---|---|---|---|---|---|---|---|
| | | glass | cpx | sp | # BSE images | *phase distribution | glass | plg | cpx | sp | # BSE images |
| B$_{100}$ | 48.02 | 48.5 (1.8) | 47.7 (1.7) | 3.8 (1.2) | 4 | H | 29.4 (21.9) | 11.6 (9.5) | 54.3 (12.5) | 4.8 (0.1) | 2 |
| B$_{80}$R$_{20}$ | 53.01 | 58.5 (2.4) | 40.6 (2.4) | 0.9 (0.2) | 4 | H | 52.9 (10.9) | 8.0 (10.0) | 38.1 (0.9) | 1.1 (0.1) | 2 |
| B$_{60}$R$_{40}$ | 57.97 | 82.6 (4.9) | 16.0 (3.8) | 1.3 (1.2) | 6 | H | 68.9 | 1.3 | 25.3 | 4.5 | 1 |
| B$_{40}$R$_{60}$ | 62.73 | 97.0 (1.0) | - | 3.0 (1.0) | 5 | H | - | - | - | - | - |
| B$_{20}$R$_{80}$ | 67.91 | - | - | - | - | - | - | - | - | - | - |
| R$_{100}$ | 73.97 | - | - | - | - | - | - | - | - | - | - |

**Table 4**

| system | phase | # point analysed | SiO$_2$ | TiO$_2$ | Al$_2$O$_3$ | Fe$_2$O$_3$ | MnO | MgO | CaO | Na$_2$O | K$_2$O | total |
|---|---|---|---|---|---|---|---|---|---|---|---|---|
| B$_{100}$ | glass | 12 | 54.33 (0.55) | 0.84 (0.16) | 19.78 (1.41) | 8.20 (0.83) | 0.14 (0.10) | 4.41 (0.89) | 9.99 (0.51) | 2.78 (0.28) | 0.06 (0.03) | 100.53 (0.52) |
| | cpx | 9 | 45.66 (0.87) | 1.16 (0.02) | 12.30 (1.12) | 9.91 (0.61) | 0.01 (0.01) | 14.08 (0.81) | 17.37 (1.51) | 0.47 (0.26) | 0.03 (0.02) | 100.92 (0.77) |
| B$_{80}$R$_{20}$ | glass | 11 | 56.56 (0.75) | 0.86 (0.08) | 18.34 (0.41) | 7.79 (0.46) | 0.11 (0.08) | 3.93 (0.58) | 8.70 (0.54) | 3.02 (0.21) | 1.21 (0.09) | 100.51 (0.38) |
| B$_{60}$R$_{40}$ | glass | 12 | 58.38 (0.53) | 0.60 (0.15) | 15.41 (0.22) | 7.91 (0.39) | 0.13 (0.08) | 5.75 (0.18) | 8.64 (0.21) | 2.38 (0.15) | 1.60 (0.10) | 100.09 (0.64) |
| B$_{40}$R$_{60}$ | glass | 9 | 63.63 (0.34) | 0.44 (0.08) | 14.92 (0.15) | 5.62 (0.26) | 0.12 (0.06) | 4.19 (0.12) | 5.94 (0.15) | 3.00 (0.11) | 2.92 (0.06) | 100.23 (0.31) |
| B$_{20}$R$_{80}$ | glass | 8 | 68.09 (0.58) | 0.30 (0.10) | 14.39 (0.57) | 4.00 (0.27) | 0.12 (0.05) | 2.38 (0.17) | 3.75 (0.09) | 3.37 (0.16) | 3.88 (0.08) | 100.28 (0.61) |
| R$_{100}$ | glass | 8 | 73.24 (0.37) | 0.13 (0.05) | 13.32 (0.23) | 2.25 (0.11) | 0.09 (0.05) | 0.50 (0.08) | 1.30 (0.06) | 3.53 (0.09) | 4.84 (0.13) | 99.20 (0.33) |

**Figure 1**

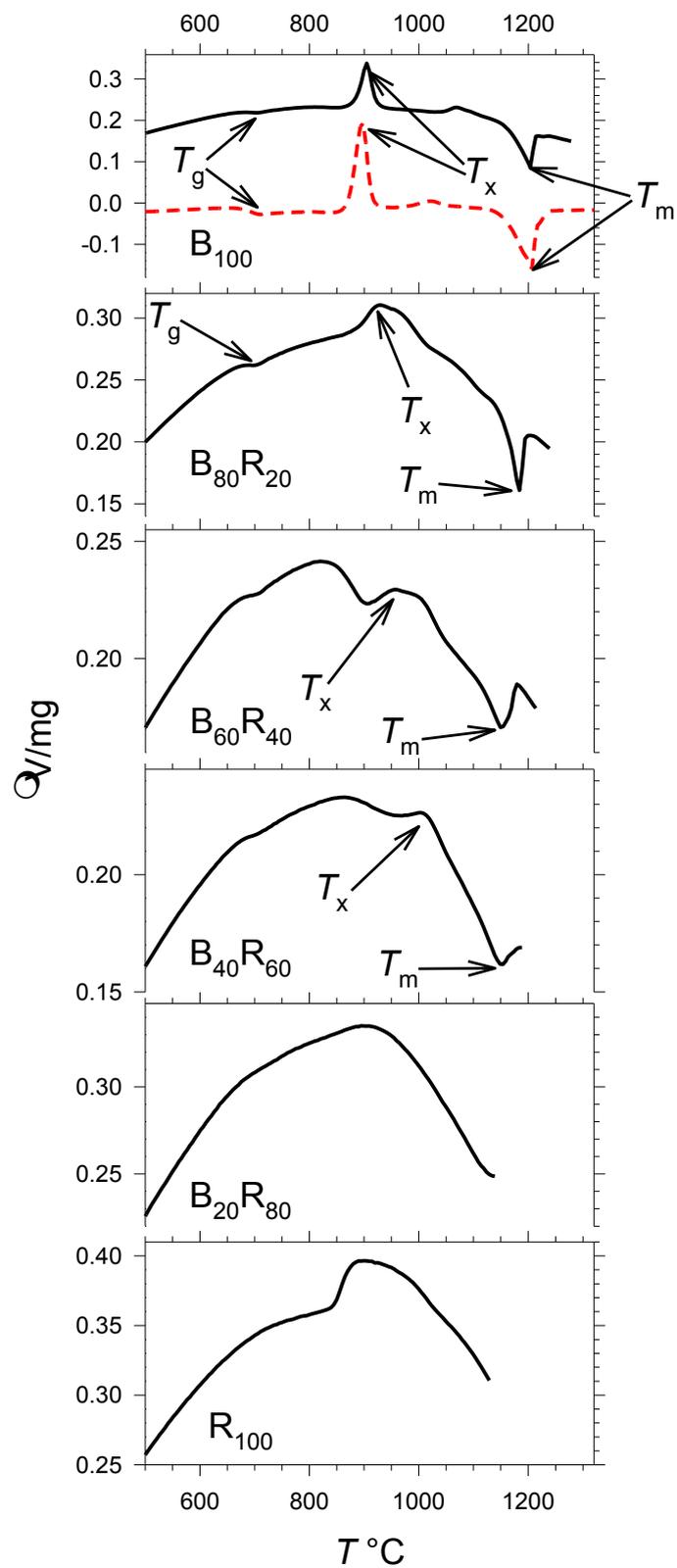

**Figure 2**

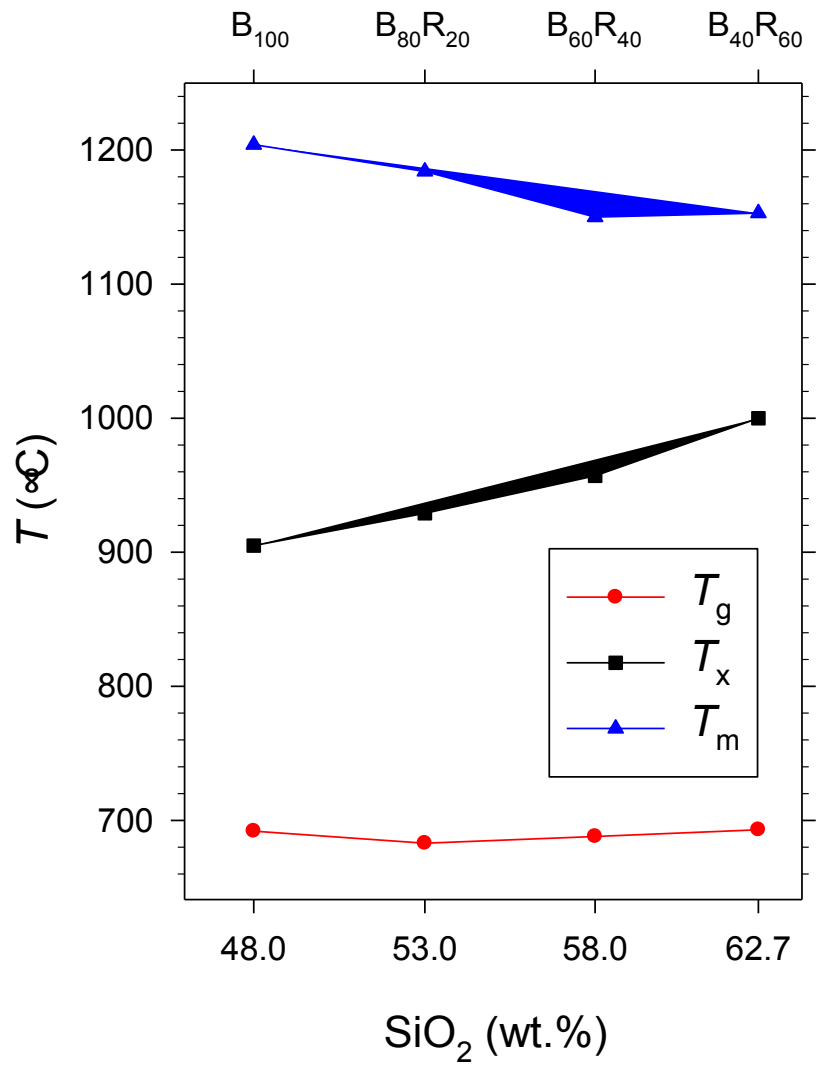

**Figure 3**

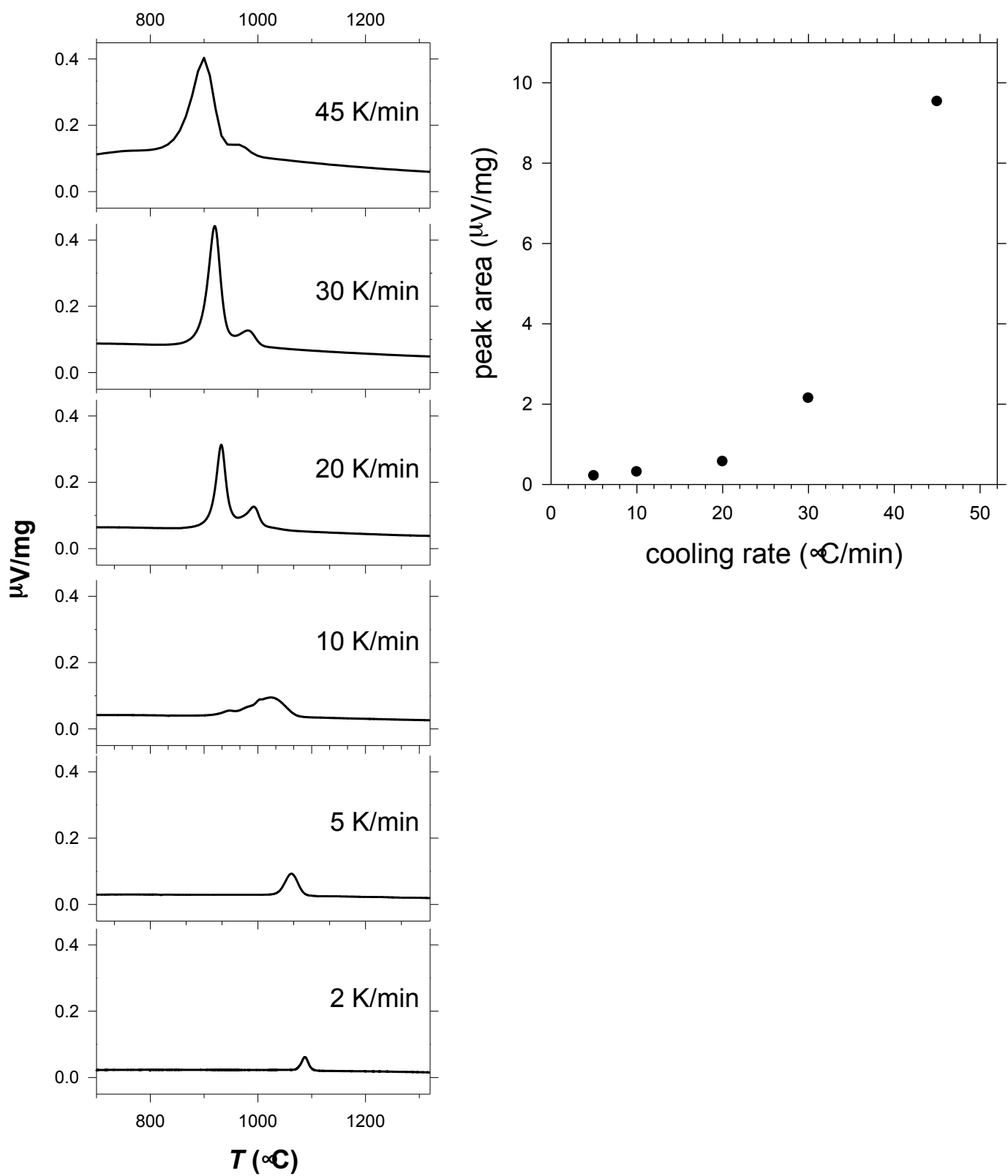

**Figure 4**

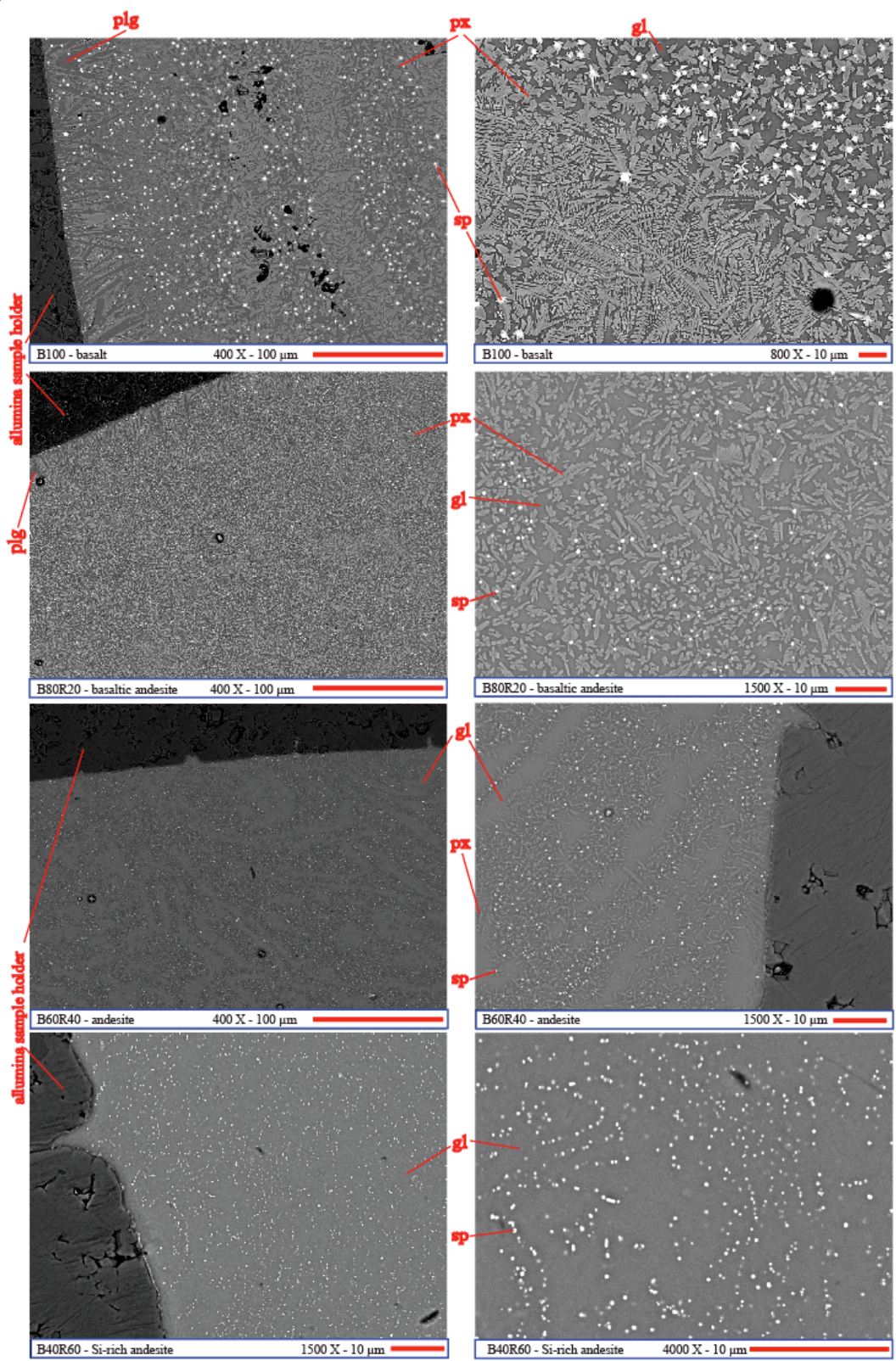

**Figure 5**

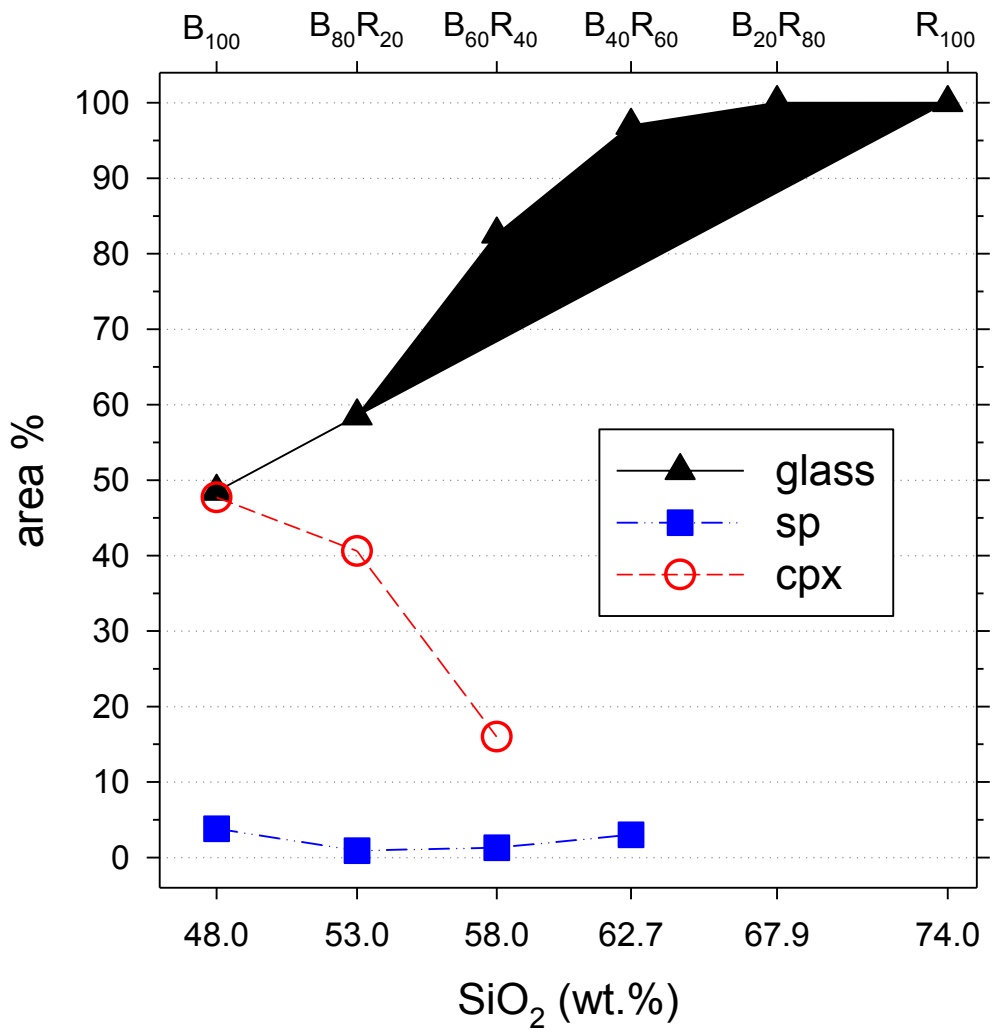

**Figure 6**

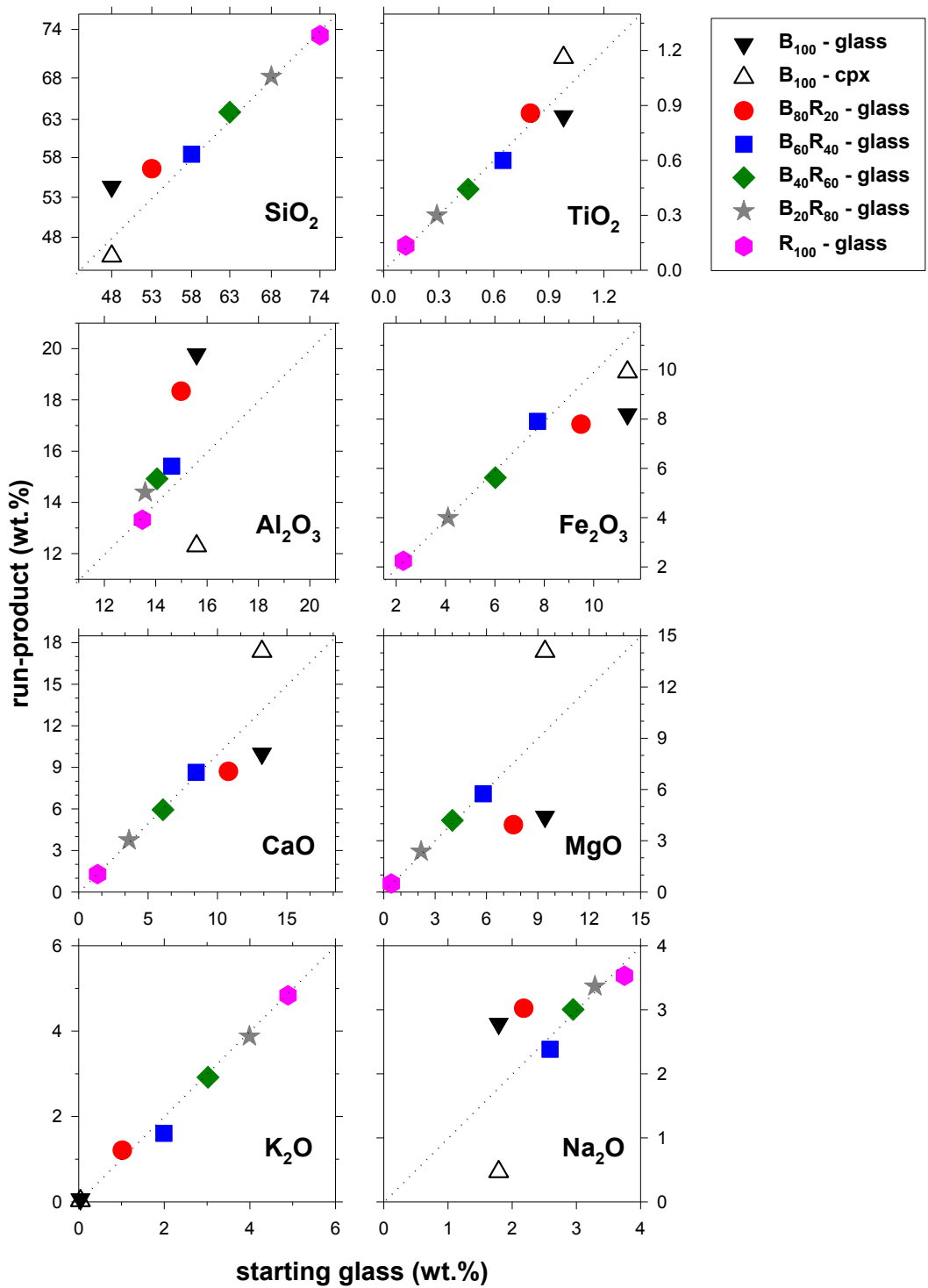

**Figure 7**

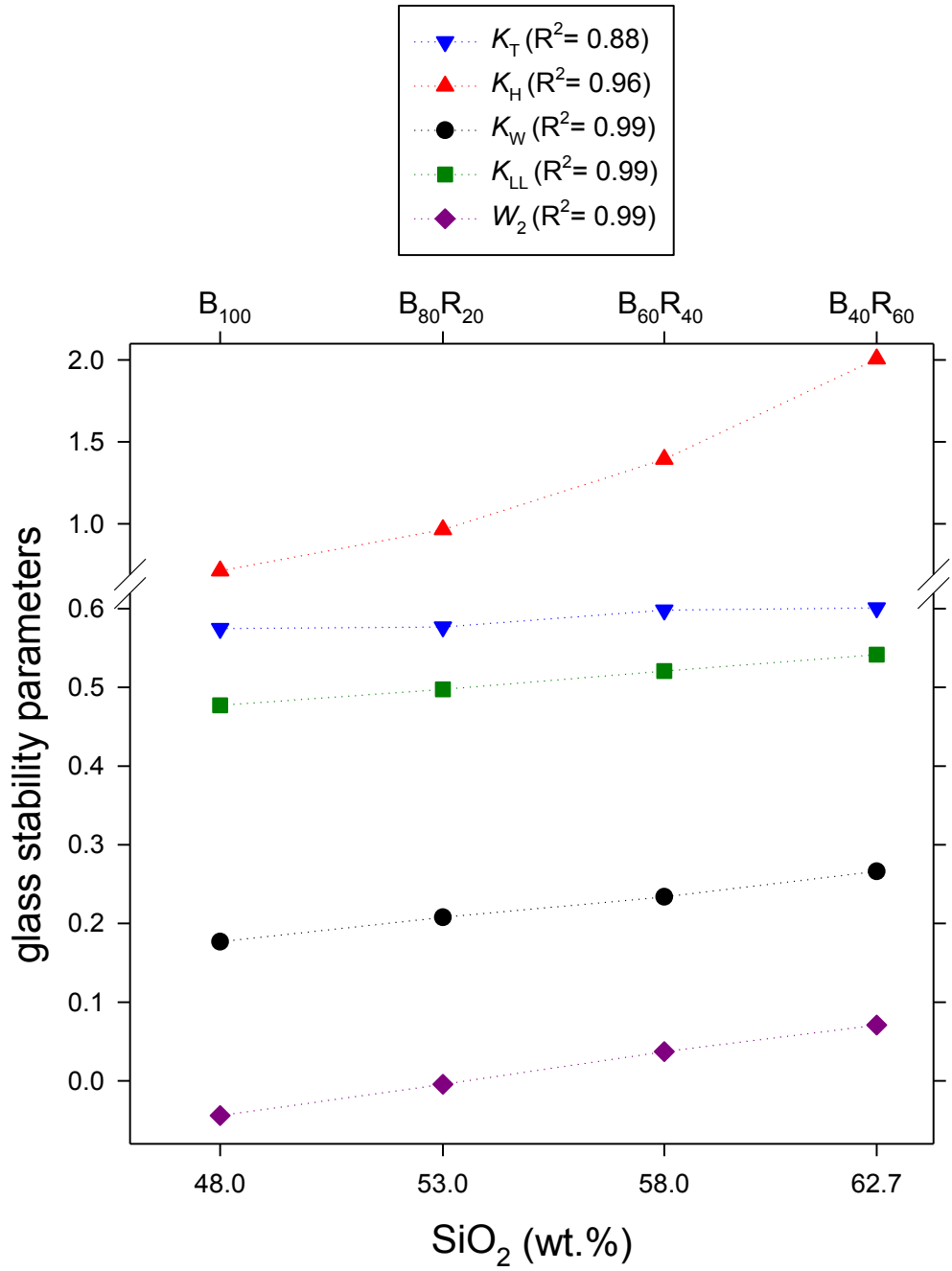

**Figure 8**

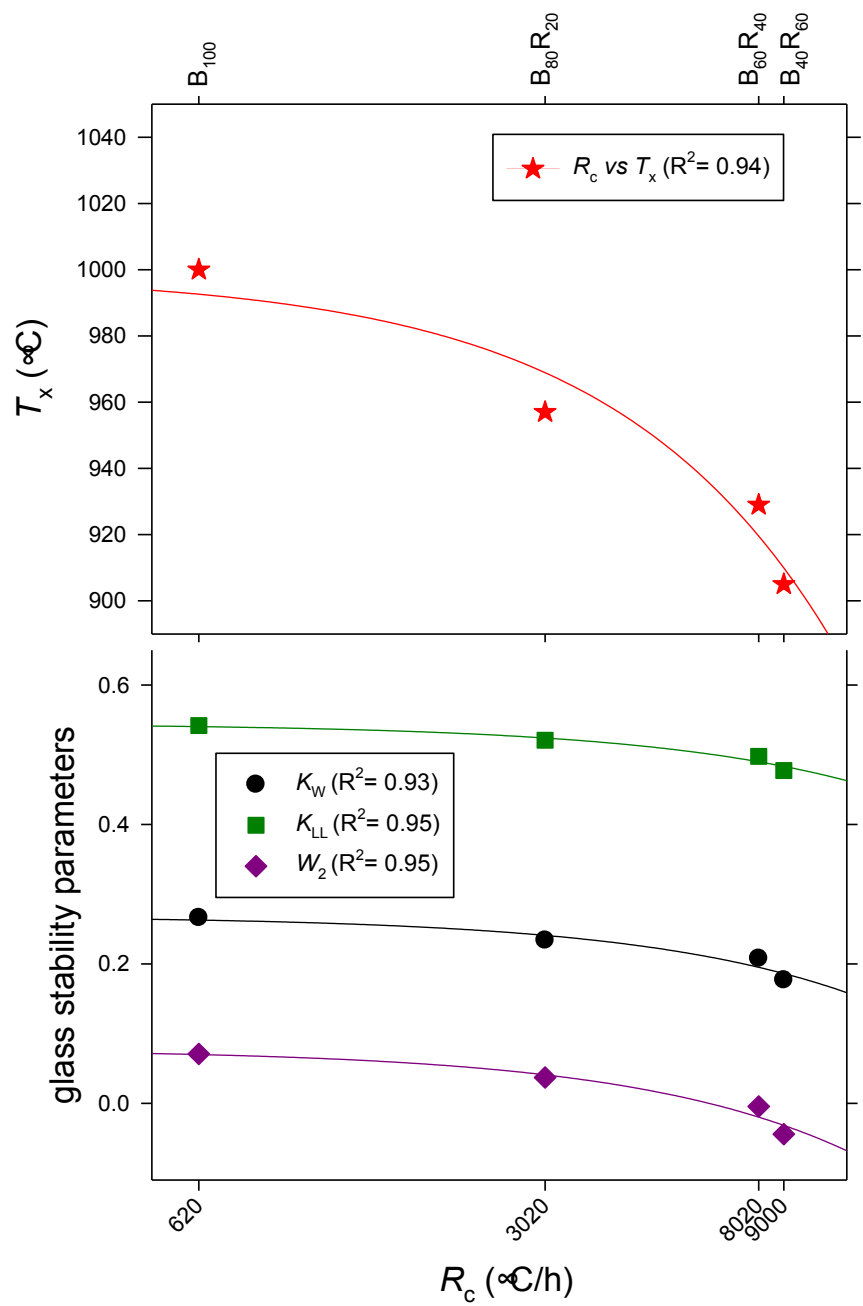